\documentclass[fleqn,10pt]{wlscirep}
\usepackage[utf8]{inputenc}
\usepackage[T1]{fontenc}
\title{qs\textit{GW} quasiparticle and \textit{GW}-BSE excitation energies of 133,885 molecules}

\author[1]{Dario Baum}
\author[1]{Arno Förster}
\author[1,*]{Lucas Visscher}
\affil[1]{Vrije Universiteit Amsterdam, Department of Chemistry and Pharmaceutical Sciences, De Boelelaan 1108, 1081 HZ Amsterdam,The Netherlands}

\affil[*]{l.visscher@vu.nl}

\begin{abstract}
Machine learning applications in the chemical sciences, especially when based on neural networks, critically depend on the availability of large quantities of high quality data. As they provide excellent accuracy for both charged and neutral excitations, a large dataset containing quasiparticle self-consistent GW (qs$GW$) and Bethe-Salpeter equation (BSE) data would be highly desirable to model excited state energies and properties. In this work, we introduce a dataset for qs$GW$-BSE excitation energies and qs$GW$ quasiparticle energies of unprecedented size. Our dataset, denoted QM9GWBSE, supplies $GW$-BSE singlet-singlet and singlet-triplet excitation energies, corresponding transition dipole moments and oscillator strengths as well as qs$GW$ quasiparticle energies for all molecules from the popular QM9 dataset. We anticipate that QM9GWBSE will provide a solid foundation to train highly accurate machine learning models for the prediction of molecular excited state properties.
\end{abstract}

\begin{document}

\flushbottom
\maketitle
%
%
\thispagestyle{empty}

\section*{Background \& Summary}

Accurate prediction and description of neutral and charged excitation energies is crucial for understanding light-matter interactions, charge transport, and spectral properties in molecular and condensed-phase systems. Computational methods enable the rational design of functional materials and provide insight into processes that are often inaccessible to direct experiment, for instance in photovoltaics \cite{photovoltaics_1} or photosynthesis \cite{photosynthesis_1, photosynthesis_2}. Rational design often requires searching for certain properties in large chemical spaces of hundreds of thousands of molecules. At the $\emph{ab initio}$ level, equation-of-motion (EOM) coupled cluster (CC) \cite{EOM-CC_1, EOM-CC_2} methods or its similarity-transformed variants (STEOM-CC)\cite{Nooijen1997,Nooijen1997a} are considered the gold standard for the calculation of excited-state properties, since they converge to the full configuration interaction (FCI) limit with increasing orders of excitation rank\cite{accuracy_EOM_CC_1, accuracy_EOM_CC_2, accuracy_EOM_CC_3}. Unfortunately, even truncated versions like EOM-CCSD (single and double excitations) EOM-CCSDT (single, double, and triple excitations) suffer from steep computational scaling of $N^6$ and $N^8$ respectively, with $N$ denoting the system size, making their use in large-scale screening studies impossible. For this reason, time-dependent \cite{TDDFT_1, TDDFT_2} density-functional theory \cite{DFT_1, DFT_2} (TD-DFT) is often the method of choice for such workflows\cite{gomez2016design, pollice2024rational}, sometimes in combination with cheaper tight-binding based approaches\cite{verma2022machine, Redox_Dyes}. However, while TD-DFT is a relatively cheap method from an \emph{ab initio} perspective, it is still computationally expensive. On top of that, it is often not accurate enough\cite{accuracy_EOM_CC_1} and suffers from an undesirable dependence on the choice of density functional. \\
Motivated by such complications, data-driven methods for the application in quantum chemical simulations have recently gained considerable attention. In particular, neural network models offer the prospect of drastically reducing the cost of electronic-structure and excited-state simulations while maintaining a level of accuracy that is often comparable to the first-principles methods which have been used for their parametrization\cite{QC_ML_1, QC_ML_2, QC_ML_3, QC_ML_5, QC_ML_6, QC_ML_7}. These approaches thus hold the promise to accelerate materials and molecular discovery without sacrificing predictive power. However, training accurate neural network models for such applications requires sufficiently large datasets of adequate data quality. Ideally, experimental data should be used for this purpose but unfortunately, experiments are usually much too costly and involved to produce sufficiently large datasets. \\
Therefore, one usually resorts to computational methods as surrogates for experimental references. TD-DFT provides an affordable alternative that could be used to generate databases of sufficient size \cite{QM9S, QCDGE}, but it is limited by its often insufficient accuracy.
Unfortunately, producing datasets of adequate size for neural network training with highly accurate wave function-based methods is likewise unfeasible due to the aforementioned computational demand. The largest and most accurate dataset of neutral excitation energies of near FCI quality is the QUEST database, consisting of roughly 1500 excitation energies of small to medium molecules\cite{accuracy_EOM_CC_1, accuracy_EOM_CC_2, veril2021questdb, loos2025quest}, which is insufficient to train neural networks. Orders of magnitude larger is the GDB-9-Ex\_EOMCCSD dataset\cite{mehta2025collection}. It contains STEOM-CCSD excitation energies of $\sim$~80\,000 molecules of the GDB-9 database with automated active space selection\cite{dutta2017automatic}. STEOM-CCSD reaches errors of only 50-200 meV in typical benchmarks for different classes of excitation energies and is usually more accurate than EOM-CCSD, whose accuracy diminishes when going from small to medium molecules\cite{Loos2020e}. \\
In search for an electronic structure method which can be used to calculate even larger databases of accurate excited state energies and properties, the $GW$ approximation (GWA) to the electronic self-energy\cite{GWA_1, GWA_2, G0W0_1, marie2024gw} combined with the Bethe-Salpeter equation (BSE) \cite{BSE_1, BSE_2, why_OP_1, why_QP_2} provides an affordable middle-ground. The GWA gives access to molecular charged excitations like ionization potentials (IP) and electron affinities (EA), and the subsequent solution of a BSE with the $GW$ quasiparticle (QP) energies and statically screened electron-hole interaction as input gives access to excited states and absorption spectra. Most $GW$ calculations neglect the off-diagonal elements of the self-energy and calculate QP energies as a perturbative correction to the Kohn--Sham (KS) eigenvalues ($G_0W_0$)\cite{G0W0_2, G0W0_3}, often combined with an iterative update of the $G_0W_0$ eigenvalues [eigenvalue self-consistent $GW$ (ev$GW$)]\cite{evGW}. Careful selection of the KS starting point, results in QP energies with deviations of only 100-200\,meV compared to highly accurate reference values in weakly correlated systems\cite{GW_IP_accurate_1, GW_IP_accurate_2, GW_IP_accurate_3, GW_IP_accurate_4, GW_IP_accurate_5, GW_IP_accurate_6}. Also for singlet-singlet neutral excitation energies, $GW$-BSE matches the accuracy of the more expensive STEOM/EOM-CCSD methods\cite{jacquemin2017bethe, gui2018accuracy, qsGW-BSE, GW_IP_accurate_4}, or even outperforms them\cite{knysh2024reference} while it is less accurate for singlet-triplet excitations\cite{Bruneval2015, Rangel2017, jacquemin2017bethe, jacquemin2017benchmark, gui2018accuracy}. \\
Despite this excellent trade-off of accuracy and computational efficiency, datasets of $GW$(-BSE) QP energies and excitation energies of sufficient size to train neural networks are rare. Examples of publicly available $GW$ datasets are the GW5000 subset of the OE62 dataset\cite{OE62} and the dataset of Fediai~\emph{et al.}\cite{SOTA_dataset}. The GW5000 subset of the OE62 dataset provides $G_0W_0$@PBE0 \cite{PBE0_1, PBE0_2, PBE0_3} quasiparticle energies for 5000 molecules with up to 100 atoms. While being especially useful for benchmarking purposes\cite{Forster2020b} or to train $\Delta$-ML models\cite{westermayr2021physically, DeltaLearning}, such data quantity usually is not enough for training neural networks for end-to-end prediction, i.e. prediction of the desired property directly from the molecular structure \cite{NNsNeedLotsData_1, NNsNeedLotsData_2, NNsNeedLotsData_3}. Exceptional in this regard is the dataset by Fediai~\emph{et al.} which contains IPs and EAs for all of the 133\,885 data points in the QM9 database\cite{QM9_1, QM9_2} calculated using ev$GW$@PBE. In a follow-up paper, they demonstrated their dataset to be sufficient for training robust and accurate DimeNet++ and SchNet models \cite{SOTA_model, SchNet_1, SchNet_2, DimeNetPlusPlus}. However, neutral excitation energies calculated with the BSE are missing as of yet.

To remedy this deficiency, we present here the largest $GW$-BSE dataset to date: we provide $GW$-BSE singlet-singlet and singlet-triplet excitation energies together with transition dipole moments and oscillator strengths for the complete QM9 dataset. Rather than using ev$GW$, we thereby decided to perform all of our $GW$ calculation in a quasi-particle self-consistent fashion (qs$GW$) which self-consistently updates both QP energies and corresponding orbitals\cite{Faleev2004, qsGW_1, qsGW_2, bruneval_springer2014}. ev$GW$ largely removes the dependence on the starting point for QP energies and frequently leads to excellent agreement with high-accuracy reference values\cite{blase2011first, faber2011first, jacquemin2015benchmarking, blase2016gw, rangel2016evaluating}, while the effect of updating also the orbitals is usually rather small\cite{faber2013many}. For neutral excitation energies, the situation is however a bit more nuanced. While for Thiel's set\cite{Schreiber2008} the mean average deviation between ev$GW$@PBE-BSE and ev$GW$@PBE0-BSE neutral excitation energies was found to be as small as 80 meV\cite{jacquemin2015benchmarking}, more recent work\cite{hashemi2021assessment, Kshirsagar2023} found a much stronger dependence of $GW$-BSE excited state energies on the KS orbitals. Adopting the only slightly more expensive\cite{LowOrderScaling_qsGW} qs$GW$ approach overcomes the dependence of QP and excited-state energies and properties on the choice of a density functional in diagonal approximations to the self-energy\cite{GWBSE_accurate, Bruneval2015, GW_IP_accurate_1, GW_IP_accurate_2}. \\
The qs$GW$ method has been shown to be among the most accurate of all $GW$ methods\cite{GW_IP_accurate_6, Forster2025} for a standard benchmark set of QP energies of 24 organic acceptor molecules\cite{richard2016accurate}. Moreover, qs$GW$-BSE provides highly reliable excited-state energies across a wide range of molecular systems\cite{gui2018accuracy, qsGW-BSE, Forster2025}. Overall, qs$GW$(-BSE) constitutes a robust, parameter-free framework for predicting accurate quasiparticle and excitonic properties, with an excellent trade-off between accuracy and computational effort. For this reason, our dataset offers an unprecedented combination of high data quantity, quality, and diversity of properties, making it an excellent choice for developing neural-network-based models for a variety of applications in molecular spectroscopy.

\section*{Methods}

All calculations were performed with the ADF engine of the Amsterdam modeling suite (AMS) \cite{AMS} via the PLAMS toolkit\cite{PLAMS}. We used the molecular structures of the QM9 dataset, optimized at the B3LYP/6-31G(2df,p) level of theory \cite{QM9_2}. For the DFT step preceding the $GW$ calculations, we used the BHANDH functional \cite{BH, LYP} as implemented in libXC\cite{Lehtola2018}. For the subsequent qs$GW$ and qs$GW$-BSE steps, we used the respective implementations in ADF \cite{Forster2020b, LowOrderScaling_qsGW, qsGW-BSE}. For both, we used the TZ3P basis set \cite{GW100_Slater}. We used minimax grids with 16 points in imaginary time and imaginary frequency, and evaluated the self-energy on the real axis through analytical continuation with a 16-point Padé approximant\cite{GW100_Slater}, following the algorithm by Vidberg and Serene\cite{Vidberg1977}.  For all qs$GW$-BSE calculations, we use the default value of 10 for the maximum number of iterations in the qs$GW$ part and the direct inversion iterative subspace (DIIS) method\cite{Pulay1980, Veril2018, LowOrderScaling_qsGW} for convergence acceleration. Further, the maximum number of iterations of the Davidson algorithm is set to 20. Based on previous experience, both values are sufficient to converge qs$GW$ and the qs$GW$-BSE calculation to a precision of a few meV\cite{qsGW-BSE}. We calculate the lowest 5 excitation energies for both singlet-singlet and singlet-triplet excitations. \\
For other numerical settings, we distinguish two cases: initially, we run qs$GW$-BSE calculations for all molecules with the numerical quality set to \textit{Good}, and eliminate almost linear dependent products of basis functions from the primary basis by setting a $K$-matrix regularisation parameter\cite{Spadetto2023} to $\epsilon_{K} = 5 \times 10^{-3}$. This entails the use of an auxiliary basis consisting of auxiliary functions with angular momentum up to $ l=4$ \cite {Forster2020} in the pair atomic density fitting (PADF)\cite{Spadetto2023} approximation to the 2-electron integrals on which our implementation is based\cite{Forster2020b}. We choose this setting as the default because it is very efficient and usually reliable\cite{LowOrderScaling_qsGW, qsGW-BSE}. However, the auxiliary basis should ideally contain auxiliary functions with angular momenta $l=5$ and $l=6$ to be able to accurately represent products involving the $f$-functions contained in the TZ3P basis set for second-row elements. The lack of these functions can hinder convergence of either the qs$GW$ calculation or the Davidson diagonalization, or, in the worst case, induce a variational collapse. We detect those rare cases using a variety of automatic filters described in Section~\textit{Technical Validation}). All calculations that trigger one of our automatic filters are restarted with $\epsilon_{K} = 10^{-3}$, and the numerical quality set to \textit{VeryGood}, entailing the use of an auxiliary basis with functions of with angular momentum up to $ l=6$ \cite {Forster2020}. With these settings, all qs$GW$-BSE calculations can be safely converged with numerical errors which are at least an order of magnitude smaller than the errors inherent in the method. For full reproducibility, we include the ADF input files for both the initial calculations and the restart calculations in the SI.

\section*{Data Records}

The dataset is available at <LINK TO BE INCLUDED SOON>. The repository contains a zip file for each property. Each zip file includes a data file for each molecule with the filenames being \texttt{mol}\_\texttt{ID} where \texttt{ID} is the same numerical molecule identifier as in the QM9 dataset with leading zeros left out (e.g. \glqq \texttt{4200}\grqq{} instead of \glqq \texttt{004200}\grqq{}). Each data file contains the numerical values of the property at hand as a 1D-array in ascending order, so from lowest to highest energy (e.g. lowest occupied to highest virtual orbital energy or lowest to highest excitation energy). Tab. \ref{tab: data legend} maps each zip file name to its respective property with the corresponding unit.

\begin{table}[ht]
    \centering
    \begin{tabular}{|l|l|l|}
        \hline
        Filename & Description & Unit \\
        \hline
        eqp & qs$GW$ quasiparticle energies & eV \\
        \hline
        eexc\_ss & qs$GW$-BSE singlet-singlet excitation energies & eV \\
        \hline
        eexc\_st & qs$GW$-BSE singlet-triplet excitation energies & eV \\
        \hline
        trans\_dip\_mom & Transition dipole moments of singlet-singlet excitations & D \\
        \hline
        osc\_stren & Oscillator strengths of singlet-singlet excitations & $-$ \\
        \hline
        edft & DFT MO energies & eV \\
        \hline
        xyz & Molecular structures & $-$ \\
        \hline
    \end{tabular}
    \caption{Overview of properties in the QM9GWBSE dataset with corresponding units and filenames in the data repository.}
    \label{tab: data legend}
\end{table}

\begin{figure}[hbt!]
    \centering
    \includegraphics[width=\textwidth]{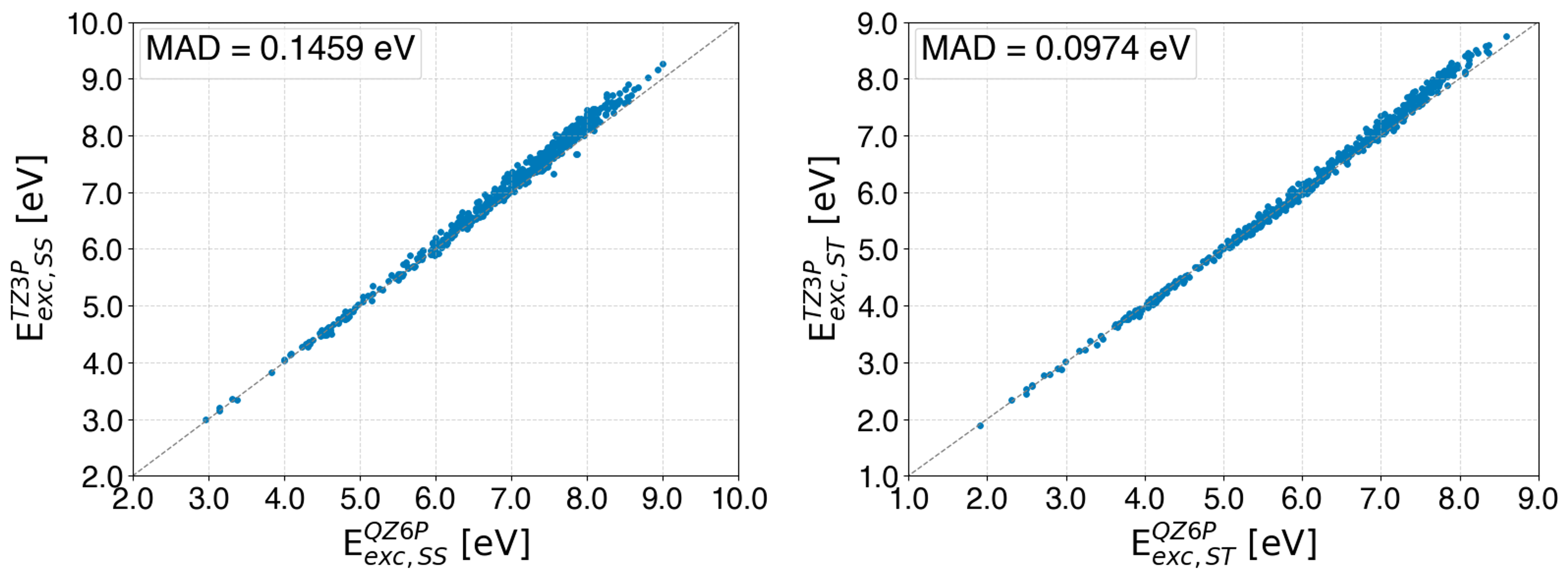}
    \caption{Deviation of $GW$-BSE singlet-singlet (left) and singlet-triplet (right) excitation energies based on the TZ3P and QZ6P basis sets.}
    \label{fig: BSE TZ3P vs QZ6P}
\end{figure}

\section*{Technical Validation}

The large amount of calculations performed in this work prevents manual inspection of all generated data. Therefore, we ensured data quality by including automated filters to check the physical plausibility of each calculation as well as statistical analysis of the data, ensuring that no outliers or systematic artifacts are present, and to compare our work to the current state of the art.

\subsection*{Automated Filters}

We used several automated checks to detect and restart calculations with unphysical results. The checks are based on the physical plausibility of the quasiparticle energies of the $GW$ calculations in relation to the MO energies of the underlying DFT calculations. First, for organic molecules, the gap between the qs$GW$ highest occupied molecular orbital (Homo) and lowest unoccupied molecular orbital (Lumo) is almost always larger than the ones of hybrid functionals like BHANDH with 50\,\% exact exchange\cite{LowOrderScaling_qsGW}. Second, the $GW$ Homo QP will be lower than the corresponding DFT MO energy, and the Lumo energy will be higher\cite{LowOrderScaling_qsGW}. Third, within the QM9 set of molecules, the Homo QP energy should not be lower than $-20$\,eV to be plausible. With these simple checks, we can detect all cases of variational collapse due to an insufficient auxiliary basis set or of convergence of the qs$GW$ calculation to an unphysical solution\cite{LowOrderScaling_qsGW}. Additionally, we checked for imaginary eigenvalues in the Davidson algorithm, which would certainly arise from a variational collapse in the qs$GW$ calculation. \\
Any calculation that triggered one or more of these filters (less than a percent) was restarted with the tight settings described in the \textit{Methods} section. For such restarted calculations, the exact same quality checks were applied. In all but two cases, restarting with tight settings led to a normal termination of the calculation, passing all filters. For both outliers, \texttt{mol\_37992} and \texttt{mol\_133858}, only the $GW$-BSE part of the calculation failed due to imaginary eigenvalues in the Davidson procedure. The prior qs$GW$ calculations converged as expected with reliable results in both cases. The outliers have a qs$GW$ QP Homo-Lumo gap of 4.532\,eV and 5.847\,eV, respectively, which is substantially lower than the mean of 11.146\,eV over the whole dataset. Additionally, both molecules display negative QP Lumo energies. We obtain equivalent results, low QP gaps with negative Lumos, for equivalent $G_0W_0$ calculations. Furthermore, TD-DFT calculations for both molecules also fail for the singlet-triplet calculations also with imaginary eigenvalues in the Davidson algorithm. All in all, this indicates the existence of a singlet-triplet instability. The two outlier cases are specified in the README file of the data repository.

\subsection*{Basis Set Convergence}

To validate the choice of the TZ3P basis set for $GW$-BSE calculations, we demonstrate that it is close to convergence with respect to the basis set quality. For this purpose, we randomly sample 100 QM9 molecules and perform $GW$-BSE calculations with the exact same settings as for our dataset but with the QZ6P basis set instead of the TZ3P basis set and with the numerical quality set to \textit{VeryGood} instead of \textit{Good}. Generally, $GW$-BSE excitation energies with QZ6P can be considered converged in terms of basis size and are therefore a suitable yet feasible target for this quality check\cite{qsGW-BSE}. We thus compare all resulting singlet-singlet and singlet-triplet test excitation energies with the respective TZ3P equivalents from our dataset. The results are plotted in Fig.\,\ref{fig: BSE TZ3P vs QZ6P}. Especially for lower excitation energies, we get excellent agreement between QZ6P and TZ3P for both singlet-singlet and singlet-triplet excitations. For higher-lying excitation energies, the deviations grows a bit larger in both cases but are still fairly small. The average deviation is of the same order of magnitude as the typical errors of accurate $GW$-BSE calculations with respect to highly accurate wave function-based reference values\cite{Loos2021, gui2018accuracy, GW_IP_accurate_4, Forster2025}. Higher excitation energies often correspond to excitations to diffuse virtual orbitals and might have substantial Rydberg character. In such cases, more diffuse basis functions are needed to relax the corresponding orbitals. That explains why, especially for these cases, in the TZ3P basis, the excitation energies are slightly overestimated compared to the results using the QZ6P. \\
We also note that, as all $GW$-BSE methods, also qs$GW$ tends to underestimate singlet-triplet excitations\cite{gui2018accuracy, Bruneval2015}. For this reason, the basis set incompleteness error in our calculations results in a favourable error cancellation. For singlet-singlet excitations, qs$GW$-BSE does not exhibit any clear trend to either over- or underestimate excitation energies. 

\begin{figure}[ht!]
    \centering
    \includegraphics[width=\textwidth]{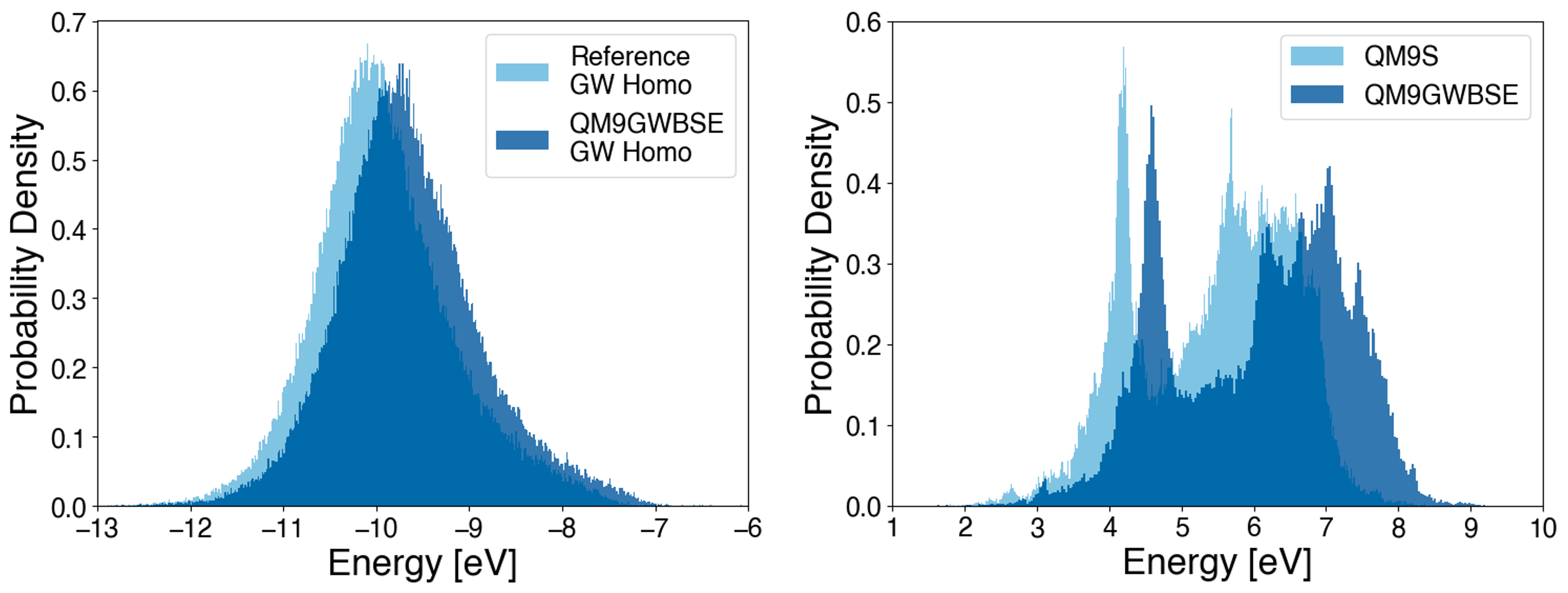}
    \caption{Distribution of GW Homo energies from the QM9GWBSE and the chosen reference dataset (left) and distribution of $GW$-BSE excitation energies from the QM9GWBSE dataset and TD-DFT excitation energies from the QM9S dataset (right).}
    \label{fig: gw and gw-bse comparison}
\end{figure}

\subsection*{Data Analysis}

We assess the overall consistency of the data by analyzing distributions of properties to confirm the absence of outliers and systematic errors. Furthermore, we compare those contributions to equivalent state-of-the-art work to demonstrate the plausibility of our results. For comparison of the QP energies, we choose the work by Fediai \emph{et al.}\cite{SOTA_dataset} who also computed $GW$ QP energies for the QM9 set of molecules, but using $G_0W_0$@PBE and ev$GW$@PBE instead of qs$GW$ as in our case. They extrapolated the resulting $GW$ QP energies to the complete basis set (CBS) limit using a commonly used two-point extrapolation scheme that presupposes dependence of the basis set incompleteness error on the inverse of number of basis functions\cite{VanSetten2013, GW_IP_accurate_6, Golze2018, Golze2020, OE62, GW100_Slater, slow_basis_set_convergence_1, GWExtrpol_1, GWExtrpol_2, GWExtrpol_3}. They performed the extrapolation using the aug-cc-DZVP and aug-cc-TZVP basis sets, which are usually too small to allow for a reliable extrapolation\cite{QP_CBS_Extrpol}. This is also apparent from the calculations by Fediai \emph{et al.}, who point out severe outliers in their CBS extrapolation of both $G_0W_0$@PBE and ev$GW$@PBE QP HOMO and LUMO energies\cite{SOTA_dataset}. \\
On top of that, recent work has demonstrated that for qs$GW$, the basis set convergence of QP energies heavily depends on the molecule at hand\cite{QP_CBS_Extrpol}. Therefore, the two-point extrapolation \cite{TwoPointExtrpol_1, TwoPointExtrpol_2} cannot be applied without introducing a considerable amount of outliers, which would result in unreliable data quality. This would even be the case if a QZ basis set had been used in the extrapolation\cite{QP_CBS_Extrpol}. Also the recently introduced data-driven extrapolation schemes, which use the orbital kinetic energy as a descriptor for basis set incompleteness\cite{Bruneval2020, QP_CBS_Extrpol} do not work well for qs$GW$\cite{QP_CBS_Extrpol}. For this reason, we did not extrapolate our QP energies to the CBS limit. \\
Fig.~\ref{fig: gw and gw-bse comparison} on the left shows the distribution of ev$GW$@PBE Homo QP energies with CBS extrapolation from Fediai et al. and the qs$GW$ QP Homo energies from our work. As can be seen, both their and our datasets agree qualitatively on the form of the distributions. On average, the qs$GW$ Homo energies in our implementation are roughly 0.25\,eV lower than the ev$GW$@PBE ones from Fediai et al. We observe that our qs$GW$ Homo energies are, on average, around 0.15\,eV lower than the ev$GW$@PBE Homo energies \textit{without} CBS extrapolation. Furthermore, we observe, again on average, that the CBS extrapolation lowers their Homo energies by roughly 0.40\,eV. Thus, the observed shift between the mean of our qs$GW$ Homo energies and the mean of the reference ev$GW$@PBE Homo energies can be explained by partial cancellation of both effects. \\
As a reference to compare our qs$GW$-BSE excitation energies, we choose the QM9S dataset\cite{QM9S}. It provides TD-DFT excitation energies on the $\omega$B97X-D/6-31G(d) level of theory for most QM9 molecules that additionally were reoptimized on the B3LYP-D3(BJ)/6-31G(d) level of theory. Fig.\,\ref{fig: gw and gw-bse comparison} on the right shows the distribution of the respective lowest singlet-singlet qs$GW$-BSE excitation energies together with the respective lowest QM9S TD-DFT excitation energies. Both distributions have the same qualitative structure, with a sharp peak within the range 3\,eV to 5\,eV and a broader group of overlapping peaks in the range 5\,eV to 9\,eV. The difference between both distributions is that our qs$GW$-BSE excitation energies are shifted towards larger energies. \\
Overall, through comparison to similar work from the literature, we can confirm that our data is free of systematic errors. Moreover, all presented data distributions demonstrate the absence of implausible outliers.

\subsection*{Code Availability}

The AMS software package containing the ADF engine and the PLAMS toolkit is available from \url{https://www.scm.com/downloads/} for a license fee. Template ADF input files used in this work can be found in the SI. Additionally, a template PLAMS script used to generate the ADF input file, run the corresponding calculation and check results is also included in the data repository. Further code is not required to reproduce the data presented here.

\bibliography{sample}

\begin{thebibliography}{100}
\urlstyle{rm}
\expandafter\ifx\csname url\endcsname\relax
  \def\url#1{\texttt{#1}}\fi
\expandafter\ifx\csname urlprefix\endcsname\relax\def\urlprefix{URL }\fi
\expandafter\ifx\csname doiprefix\endcsname\relax\def\doiprefix{DOI: }\fi
\providecommand{\bibinfo}[2]{#2}
\providecommand{\eprint}[2][]{\url{#2}}

\bibitem{photovoltaics_1}
\bibinfo{author}{Gruber, M.} \emph{et~al.}
\newblock \bibinfo{journal}{\bibinfo{title}{Thermodynamic efficiency limit of molecular donor-acceptor solar cells and its application to diindenoperylene/c60-based planar heterojunction devices}}.
\newblock {\emph{\JournalTitle{Advanced Energy Materials}}} \textbf{\bibinfo{volume}{2}}, \bibinfo{pages}{1100--1108} (\bibinfo{year}{2012}).

\bibitem{photosynthesis_1}
\bibinfo{author}{Cheng, Y.-C.} \& \bibinfo{author}{Fleming, G.~R.}
\newblock \bibinfo{journal}{\bibinfo{title}{Dynamics of light harvesting in photosynthesis}}.
\newblock {\emph{\JournalTitle{Annual review of physical chemistry}}} \textbf{\bibinfo{volume}{60}}, \bibinfo{pages}{241--262} (\bibinfo{year}{2009}).

\bibitem{photosynthesis_2}
\bibinfo{author}{Curutchet~Barat, C.~E.} \& \bibinfo{author}{Mennucci, B.}
\newblock \bibinfo{journal}{\bibinfo{title}{Quantum chemical studies of light harvesting}}.
\newblock {\emph{\JournalTitle{Chemical Reviews, 2017, vol. 117, num. 2, p. 294-343}}}  (\bibinfo{year}{2017}).

\bibitem{EOM-CC_1}
\bibinfo{author}{Monkhorst, H.~J.}
\newblock \bibinfo{journal}{\bibinfo{title}{Calculation of properties with the coupled-cluster method}}.
\newblock {\emph{\JournalTitle{International Journal of Quantum Chemistry}}} \textbf{\bibinfo{volume}{12}}, \bibinfo{pages}{421--432} (\bibinfo{year}{1977}).

\bibitem{EOM-CC_2}
\bibinfo{author}{Stanton, J.~F.} \& \bibinfo{author}{Bartlett, R.~J.}
\newblock \bibinfo{journal}{\bibinfo{title}{The equation of motion coupled-cluster method. a systematic biorthogonal approach to molecular excitation energies, transition probabilities, and excited state properties}}.
\newblock {\emph{\JournalTitle{The Journal of chemical physics}}} \textbf{\bibinfo{volume}{98}}, \bibinfo{pages}{7029--7039} (\bibinfo{year}{1993}).

\bibitem{Nooijen1997}
\bibinfo{author}{Nooijen, M.} \& \bibinfo{author}{Bartlett, R.~J.}
\newblock \bibinfo{journal}{\bibinfo{title}{A new method for excited states: Similarity transformed equation-of-motion coupled-cluster theory}}.
\newblock {\emph{\JournalTitle{Journal of Chemical Physics}}} \textbf{\bibinfo{volume}{106}}, \bibinfo{pages}{6441--6448}, \doiprefix\url{10.1063/1.474000} (\bibinfo{year}{1997}).

\bibitem{Nooijen1997a}
\bibinfo{author}{Nooijen, M.} \& \bibinfo{author}{Bartlett, R.~J.}
\newblock \bibinfo{journal}{\bibinfo{title}{Similarity transformed equation-of-motion coupled-cluster theory: Details, examples, and comparisons}}.
\newblock {\emph{\JournalTitle{Journal of Chemical Physics}}} \textbf{\bibinfo{volume}{107}}, \bibinfo{pages}{6812--6830}, \doiprefix\url{10.1063/1.474922} (\bibinfo{year}{1997}).

\bibitem{accuracy_EOM_CC_1}
\bibinfo{author}{Loos, P.-F.} \emph{et~al.}
\newblock \bibinfo{journal}{\bibinfo{title}{A mountaineering strategy to excited states: Highly accurate reference energies and benchmarks}}.
\newblock {\emph{\JournalTitle{Journal of chemical theory and computation}}} \textbf{\bibinfo{volume}{14}}, \bibinfo{pages}{4360--4379} (\bibinfo{year}{2018}).

\bibitem{accuracy_EOM_CC_2}
\bibinfo{author}{Loos, P.-F.}, \bibinfo{author}{Scemama, A.} \& \bibinfo{author}{Jacquemin, D.}
\newblock \bibinfo{journal}{\bibinfo{title}{The quest for highly accurate excitation energies: A computational perspective}}.
\newblock {\emph{\JournalTitle{The journal of physical chemistry letters}}} \textbf{\bibinfo{volume}{11}}, \bibinfo{pages}{2374--2383} (\bibinfo{year}{2020}).

\bibitem{accuracy_EOM_CC_3}
\bibinfo{author}{Marie, A.} \& \bibinfo{author}{Loos, P.-F.}
\newblock \bibinfo{journal}{\bibinfo{title}{Reference energies for valence ionizations and satellite transitions}}.
\newblock {\emph{\JournalTitle{Journal of Chemical Theory and Computation}}} \textbf{\bibinfo{volume}{20}}, \bibinfo{pages}{4751--4777} (\bibinfo{year}{2024}).

\bibitem{TDDFT_1}
\bibinfo{author}{Runge, E.} \& \bibinfo{author}{Gross, E.~K.}
\newblock \bibinfo{journal}{\bibinfo{title}{Density-functional theory for time-dependent systems}}.
\newblock {\emph{\JournalTitle{Physical review letters}}} \textbf{\bibinfo{volume}{52}}, \bibinfo{pages}{997} (\bibinfo{year}{1984}).

\bibitem{TDDFT_2}
\bibinfo{author}{Petersilka, M.}, \bibinfo{author}{Gossmann, U.} \& \bibinfo{author}{Gross, E.}
\newblock \bibinfo{journal}{\bibinfo{title}{Excitation energies from time-dependent density-functional theory}}.
\newblock {\emph{\JournalTitle{Physical review letters}}} \textbf{\bibinfo{volume}{76}}, \bibinfo{pages}{1212} (\bibinfo{year}{1996}).

\bibitem{DFT_1}
\bibinfo{author}{Hohenberg, P.} \& \bibinfo{author}{Kohn, W.}
\newblock \bibinfo{journal}{\bibinfo{title}{Inhomogeneous electron gas}}.
\newblock {\emph{\JournalTitle{Physical Review}}} \textbf{\bibinfo{volume}{136}}, \bibinfo{pages}{B864--B871} (\bibinfo{year}{1964}).

\bibitem{DFT_2}
\bibinfo{author}{Kohn, W.} \& \bibinfo{author}{Sham, L.~J.}
\newblock \bibinfo{journal}{\bibinfo{title}{Self-consistent equations including exchange and correlation effects}}.
\newblock {\emph{\JournalTitle{Physical Review}}} \textbf{\bibinfo{volume}{140}}, \bibinfo{pages}{A1133--A1138} (\bibinfo{year}{1965}).

\bibitem{gomez2016design}
\bibinfo{author}{G{\'o}mez-Bombarelli, R.} \emph{et~al.}
\newblock \bibinfo{journal}{\bibinfo{title}{Design of efficient molecular organic light-emitting diodes by a high-throughput virtual screening and experimental approach}}.
\newblock {\emph{\JournalTitle{Nature materials}}} \textbf{\bibinfo{volume}{15}}, \bibinfo{pages}{1120--1127} (\bibinfo{year}{2016}).

\bibitem{pollice2024rational}
\bibinfo{author}{Pollice, R.}, \bibinfo{author}{Ding, B.} \& \bibinfo{author}{Aspuru-Guzik, A.}
\newblock \bibinfo{journal}{\bibinfo{title}{Rational design of organic molecules with inverted gaps between the first excited singlet and triplet}}.
\newblock {\emph{\JournalTitle{Matter}}} \textbf{\bibinfo{volume}{7}}, \bibinfo{pages}{1161--1186} (\bibinfo{year}{2024}).

\bibitem{verma2022machine}
\bibinfo{author}{Verma, S.}, \bibinfo{author}{Rivera, M.}, \bibinfo{author}{Scanlon, D.~O.} \& \bibinfo{author}{Walsh, A.}
\newblock \bibinfo{journal}{\bibinfo{title}{Machine learned calibrations to high-throughput molecular excited state calculations}}.
\newblock {\emph{\JournalTitle{The Journal of Chemical Physics}}} \textbf{\bibinfo{volume}{156}} (\bibinfo{year}{2022}).

\bibitem{Redox_Dyes}
\bibinfo{author}{Beli{\'c}, J.}, \bibinfo{author}{F{\"o}rster, A.}, \bibinfo{author}{Menzel, J.~P.}, \bibinfo{author}{Buda, F.} \& \bibinfo{author}{Visscher, L.}
\newblock \bibinfo{journal}{\bibinfo{title}{Automated assessment of redox potentials for dyes in dye-sensitized photoelectrochemical cells}}.
\newblock {\emph{\JournalTitle{Physical Chemistry Chemical Physics}}} \textbf{\bibinfo{volume}{24}}, \bibinfo{pages}{197--210} (\bibinfo{year}{2022}).

\bibitem{QC_ML_1}
\bibinfo{author}{Westermayr, J.} \& \bibinfo{author}{Marquetand, P.}
\newblock \bibinfo{journal}{\bibinfo{title}{Machine learning for electronically excited states of molecules}}.
\newblock {\emph{\JournalTitle{Chemical Reviews}}} \textbf{\bibinfo{volume}{121}}, \bibinfo{pages}{9873--9926} (\bibinfo{year}{2020}).

\bibitem{QC_ML_2}
\bibinfo{author}{Dral, P.~O.} \& \bibinfo{author}{Barbatti, M.}
\newblock \bibinfo{journal}{\bibinfo{title}{Molecular excited states through a machine learning lens}}.
\newblock {\emph{\JournalTitle{Nature Reviews Chemistry}}} \textbf{\bibinfo{volume}{5}}, \bibinfo{pages}{388--405} (\bibinfo{year}{2021}).

\bibitem{QC_ML_3}
\bibinfo{author}{Cignoni, E.} \emph{et~al.}
\newblock \bibinfo{journal}{\bibinfo{title}{Electronic excited states from physically constrained machine learning}}.
\newblock {\emph{\JournalTitle{ACS Central Science}}} \textbf{\bibinfo{volume}{10}}, \bibinfo{pages}{637--648} (\bibinfo{year}{2024}).

\bibitem{QC_ML_5}
\bibinfo{author}{Grunert, M.}, \bibinfo{author}{Gro{\ss}mann, M.} \& \bibinfo{author}{Runge, E.}
\newblock \bibinfo{journal}{\bibinfo{title}{Machine learning climbs the jacob’s ladder of optoelectronic properties}}.
\newblock {\emph{\JournalTitle{Nature Communications}}} \textbf{\bibinfo{volume}{16}}, \bibinfo{pages}{8142} (\bibinfo{year}{2025}).

\bibitem{QC_ML_6}
\bibinfo{author}{Grunert, M.}, \bibinfo{author}{Gro{\ss}mann, M.} \& \bibinfo{author}{Runge, E.}
\newblock \bibinfo{journal}{\bibinfo{title}{Discovery of sustainable energy materials via the machine-learned material space}}.
\newblock {\emph{\JournalTitle{Small}}} \bibinfo{pages}{2412519} (\bibinfo{year}{2025}).

\bibitem{QC_ML_7}
\bibinfo{author}{Grunert, M.}, \bibinfo{author}{Gro{\ss}mann, M.} \& \bibinfo{author}{Runge, E.}
\newblock \bibinfo{journal}{\bibinfo{title}{Deep learning of spectra: Predicting the dielectric function of semiconductors}}.
\newblock {\emph{\JournalTitle{Physical review materials}}} \textbf{\bibinfo{volume}{8}}, \bibinfo{pages}{L122201} (\bibinfo{year}{2024}).

\bibitem{QM9S}
\bibinfo{author}{Zou, Z.} \emph{et~al.}
\newblock \bibinfo{journal}{\bibinfo{title}{A deep learning model for predicting selected organic molecular spectra}}.
\newblock {\emph{\JournalTitle{Nature Computational Science}}} \textbf{\bibinfo{volume}{3}}, \bibinfo{pages}{957--964} (\bibinfo{year}{2023}).

\bibitem{QCDGE}
\bibinfo{author}{Zhu, Y.}, \bibinfo{author}{Li, M.}, \bibinfo{author}{Xu, C.} \& \bibinfo{author}{Lan, Z.}
\newblock \bibinfo{journal}{\bibinfo{title}{Quantum chemistry dataset with ground-and excited-state properties of 450 kilo molecules}}.
\newblock {\emph{\JournalTitle{Scientific Data}}} \textbf{\bibinfo{volume}{11}}, \bibinfo{pages}{948} (\bibinfo{year}{2024}).

\bibitem{veril2021questdb}
\bibinfo{author}{V{\'e}ril, M.} \emph{et~al.}
\newblock \bibinfo{journal}{\bibinfo{title}{Questdb: A database of highly accurate excitation energies for the electronic structure community}}.
\newblock {\emph{\JournalTitle{Wiley Interdisciplinary Reviews: Computational Molecular Science}}} \textbf{\bibinfo{volume}{11}}, \bibinfo{pages}{e1517} (\bibinfo{year}{2021}).

\bibitem{loos2025quest}
\bibinfo{author}{Loos, P.-F.}, \bibinfo{author}{Boggio-Pasqua, M.}, \bibinfo{author}{Blondel, A.}, \bibinfo{author}{Lipparini, F.} \& \bibinfo{author}{Jacquemin, D.}
\newblock \bibinfo{journal}{\bibinfo{title}{Quest database of highly-accurate excitation energies}}.
\newblock {\emph{\JournalTitle{Journal of Chemical Theory and Computation}}} \textbf{\bibinfo{volume}{21}}, \bibinfo{pages}{8010--8033} (\bibinfo{year}{2025}).

\bibitem{mehta2025collection}
\bibinfo{author}{Mehta, K.}, \bibinfo{author}{Pasini, M.~L.}, \bibinfo{author}{Ganyushin, D.}, \bibinfo{author}{Yoo, P.} \& \bibinfo{author}{Irle, S.}
\newblock \bibinfo{journal}{\bibinfo{title}{Collection: Td-dft and eom-ccsd calculations for the gdb-9-ex dataset}}.
\newblock {\emph{\JournalTitle{IEEE Data Descriptions}}}  (\bibinfo{year}{2025}).

\bibitem{dutta2017automatic}
\bibinfo{author}{Dutta, A.~K.}, \bibinfo{author}{Nooijen, M.}, \bibinfo{author}{Neese, F.} \& \bibinfo{author}{Izs{\'a}k, R.}
\newblock \bibinfo{journal}{\bibinfo{title}{Automatic active space selection for the similarity transformed equations of motion coupled cluster method}}.
\newblock {\emph{\JournalTitle{The Journal of Chemical Physics}}} \textbf{\bibinfo{volume}{146}} (\bibinfo{year}{2017}).

\bibitem{Loos2020e}
\bibinfo{author}{Loos, P.~F.}, \bibinfo{author}{Lipparini, F.}, \bibinfo{author}{Boggio-Pasqua, M.}, \bibinfo{author}{Scemama, A.} \& \bibinfo{author}{Jacquemin, D.}
\newblock \bibinfo{journal}{\bibinfo{title}{A mountaineering strategy to excited states: Highly accurate energies and benchmarks for medium sized molecules}}.
\newblock {\emph{\JournalTitle{Journal of Chemical Theory and Computation}}} \textbf{\bibinfo{volume}{16}}, \bibinfo{pages}{1711--1741}, \doiprefix\url{10.1021/acs.jctc.9b01216} (\bibinfo{year}{2020}).

\bibitem{GWA_1}
\bibinfo{author}{Hedin, L.}
\newblock \bibinfo{journal}{\bibinfo{title}{New method for calculating the one-particle green's function with application to the electron-gas problem}}.
\newblock {\emph{\JournalTitle{Physical Review}}} \textbf{\bibinfo{volume}{139}}, \bibinfo{pages}{A796} (\bibinfo{year}{1965}).

\bibitem{GWA_2}
\bibinfo{author}{Aryasetiawan, F.} \& \bibinfo{author}{Gunnarsson, O.}
\newblock \bibinfo{journal}{\bibinfo{title}{The gw method}}.
\newblock {\emph{\JournalTitle{Reports on progress in Physics}}} \textbf{\bibinfo{volume}{61}}, \bibinfo{pages}{237} (\bibinfo{year}{1998}).

\bibitem{G0W0_1}
\bibinfo{author}{Golze, D.}, \bibinfo{author}{Dvorak, M.} \& \bibinfo{author}{Rinke, P.}
\newblock \bibinfo{journal}{\bibinfo{title}{The gw compendium: A practical guide to theoretical photoemission spectroscopy}}.
\newblock {\emph{\JournalTitle{Frontiers in chemistry}}} \textbf{\bibinfo{volume}{7}}, \bibinfo{pages}{377} (\bibinfo{year}{2019}).

\bibitem{marie2024gw}
\bibinfo{author}{Marie, A.}, \bibinfo{author}{Ammar, A.} \& \bibinfo{author}{Loos, P.-F.}
\newblock \bibinfo{title}{The gw approximation: A quantum chemistry perspective}.
\newblock In \emph{\bibinfo{booktitle}{Advances in Quantum Chemistry}}, vol.~\bibinfo{volume}{90}, \bibinfo{pages}{157--184} (\bibinfo{publisher}{Elsevier}, \bibinfo{year}{2024}).

\bibitem{BSE_1}
\bibinfo{author}{Salpeter, E.~E.} \& \bibinfo{author}{Bethe, H.~A.}
\newblock \bibinfo{journal}{\bibinfo{title}{A relativistic equation for bound-state problems}}.
\newblock {\emph{\JournalTitle{Physical Review}}} \textbf{\bibinfo{volume}{84}}, \bibinfo{pages}{1232} (\bibinfo{year}{1951}).

\bibitem{BSE_2}
\bibinfo{author}{Gell-Mann, M.} \& \bibinfo{author}{Low, F.}
\newblock \bibinfo{journal}{\bibinfo{title}{Bound states in quantum field theory}}.
\newblock {\emph{\JournalTitle{Physical Review}}} \textbf{\bibinfo{volume}{84}}, \bibinfo{pages}{350} (\bibinfo{year}{1951}).

\bibitem{why_OP_1}
\bibinfo{author}{Onida, G.}, \bibinfo{author}{Reining, L.} \& \bibinfo{author}{Rubio, A.}
\newblock \bibinfo{journal}{\bibinfo{title}{Electronic excitations: density-functional versus many-body green's-function approaches}}.
\newblock {\emph{\JournalTitle{Rev. Mod. Phys.}}} \textbf{\bibinfo{volume}{74}}, \bibinfo{pages}{601--659}, \doiprefix\url{10.1103/RevModPhys.74.601} (\bibinfo{year}{2002}).

\bibitem{why_QP_2}
\bibinfo{author}{Rohlfing, M.} \& \bibinfo{author}{Louie, S.~G.}
\newblock \bibinfo{journal}{\bibinfo{title}{Electron-hole excitations and optical spectra from first principles}}.
\newblock {\emph{\JournalTitle{Phys. Rev. B}}} \textbf{\bibinfo{volume}{62}}, \bibinfo{pages}{4927--4944}, \doiprefix\url{10.1103/PhysRevB.62.4927} (\bibinfo{year}{2000}).

\bibitem{G0W0_2}
\bibinfo{author}{Hybertsen, M.~S.} \& \bibinfo{author}{Louie, S.~G.}
\newblock \bibinfo{journal}{\bibinfo{title}{First-principles theory of quasiparticles: calculation of band gaps in semiconductors and insulators}}.
\newblock {\emph{\JournalTitle{Physical review letters}}} \textbf{\bibinfo{volume}{55}}, \bibinfo{pages}{1418} (\bibinfo{year}{1985}).

\bibitem{G0W0_3}
\bibinfo{author}{Hybertsen, M.~S.} \& \bibinfo{author}{Louie, S.~G.}
\newblock \bibinfo{journal}{\bibinfo{title}{Electron correlation in semiconductors and insulators: Band gaps and quasiparticle energies}}.
\newblock {\emph{\JournalTitle{Physical Review B}}} \textbf{\bibinfo{volume}{34}}, \bibinfo{pages}{5390} (\bibinfo{year}{1986}).

\bibitem{evGW}
\bibinfo{author}{Blase, X.}, \bibinfo{author}{Attaccalite, C.} \& \bibinfo{author}{Olevano, V.}
\newblock \bibinfo{journal}{\bibinfo{title}{First-principles gw calculations for fullerenes, porphyrins, phtalocyanine, and other molecules of interest for organic photovoltaic applications}}.
\newblock {\emph{\JournalTitle{Physical Review B-Condensed Matter and Materials Physics}}} \textbf{\bibinfo{volume}{83}}, \bibinfo{pages}{115103} (\bibinfo{year}{2011}).

\bibitem{GW_IP_accurate_1}
\bibinfo{author}{Knight, J.~W.} \emph{et~al.}
\newblock \bibinfo{journal}{\bibinfo{title}{Accurate ionization potentials and electron affinities of acceptor molecules {III}: A benchmark of {GW} methods}}.
\newblock {\emph{\JournalTitle{Journal of Chemical Theory and Computation}}} \textbf{\bibinfo{volume}{12}}, \bibinfo{pages}{615--626} (\bibinfo{year}{2016}).

\bibitem{GW_IP_accurate_2}
\bibinfo{author}{Caruso, F.}, \bibinfo{author}{Dauth, M.}, \bibinfo{author}{van Setten, M.~J.} \& \bibinfo{author}{Rinke, P.}
\newblock \bibinfo{journal}{\bibinfo{title}{Benchmark of {GW} approaches for the {GW} 100 test set}}.
\newblock {\emph{\JournalTitle{Journal of Chemical Theory and Computation}}} \textbf{\bibinfo{volume}{12}}, \bibinfo{pages}{5076--5087} (\bibinfo{year}{2016}).

\bibitem{GW_IP_accurate_3}
\bibinfo{author}{Bruneval, F.}, \bibinfo{author}{Dattani, N.} \& \bibinfo{author}{van Setten, M.~J.}
\newblock \bibinfo{journal}{\bibinfo{title}{The {GW} miracle in many-body perturbation theory for the ionization potential of molecules}}.
\newblock {\emph{\JournalTitle{Frontiers in Chemistry}}} \textbf{\bibinfo{volume}{9}}, \bibinfo{pages}{749779} (\bibinfo{year}{2021}).

\bibitem{GW_IP_accurate_4}
\bibinfo{author}{McKeon, C.~A.}, \bibinfo{author}{Hamed, S.~M.}, \bibinfo{author}{Bruneval, F.} \& \bibinfo{author}{Neaton, J.~B.}
\newblock \bibinfo{journal}{\bibinfo{title}{An optimally tuned range-separated hybrid starting point for \textit{ab initio} {GW} plus bethe--salpeter equation calculations of molecules}}.
\newblock {\emph{\JournalTitle{The Journal of Chemical Physics}}} \textbf{\bibinfo{volume}{157}}, \bibinfo{pages}{071101} (\bibinfo{year}{2022}).

\bibitem{GW_IP_accurate_5}
\bibinfo{author}{Bruneval, F.} \& \bibinfo{author}{F{\"o}rster, A.}
\newblock \bibinfo{journal}{\bibinfo{title}{Fully dynamic {G3W2} self-energy for finite systems: Formulas and benchmark}}.
\newblock {\emph{\JournalTitle{Journal of Chemical Theory and Computation}}} \textbf{\bibinfo{volume}{20}}, \bibinfo{pages}{3218--3230} (\bibinfo{year}{2024}).

\bibitem{GW_IP_accurate_6}
\bibinfo{author}{F{\"o}rster, A.} \& \bibinfo{author}{Visscher, L.}
\newblock \bibinfo{journal}{\bibinfo{title}{Exploring the statically screened {G3W2} correction to the {GW} self-energy: Charged excitations and total energies of finite systems}}.
\newblock {\emph{\JournalTitle{Physical Review B}}} \textbf{\bibinfo{volume}{105}}, \bibinfo{pages}{125121} (\bibinfo{year}{2022}).

\bibitem{jacquemin2017bethe}
\bibinfo{author}{Jacquemin, D.}, \bibinfo{author}{Duchemin, I.} \& \bibinfo{author}{Blase, X.}
\newblock \bibinfo{journal}{\bibinfo{title}{Is the bethe--salpeter formalism accurate for excitation energies? comparisons with td-dft, caspt2, and eom-ccsd}}.
\newblock {\emph{\JournalTitle{The journal of physical chemistry letters}}} \textbf{\bibinfo{volume}{8}}, \bibinfo{pages}{1524--1529} (\bibinfo{year}{2017}).

\bibitem{gui2018accuracy}
\bibinfo{author}{Gui, X.}, \bibinfo{author}{Holzer, C.} \& \bibinfo{author}{Klopper, W.}
\newblock \bibinfo{journal}{\bibinfo{title}{Accuracy assessment of {GW} starting points for calculating molecular excitation energies using the bethe--salpeter formalism}}.
\newblock {\emph{\JournalTitle{Journal of Chemical Theory and Computation}}} \textbf{\bibinfo{volume}{14}}, \bibinfo{pages}{2127--2136} (\bibinfo{year}{2018}).

\bibitem{qsGW-BSE}
\bibinfo{author}{F{\"o}rster, A.} \& \bibinfo{author}{Visscher, L.}
\newblock \bibinfo{journal}{\bibinfo{title}{Quasiparticle self-consistent gw-bethe--salpeter equation calculations for large chromophoric systems}}.
\newblock {\emph{\JournalTitle{Journal of chemical theory and computation}}} \textbf{\bibinfo{volume}{18}}, \bibinfo{pages}{6779--6793} (\bibinfo{year}{2022}).

\bibitem{knysh2024reference}
\bibinfo{author}{Knysh, I.} \emph{et~al.}
\newblock \bibinfo{journal}{\bibinfo{title}{Reference cc3 excitation energies for organic chromophores: Benchmarking td-dft, bse/gw, and wave function methods}}.
\newblock {\emph{\JournalTitle{Journal of Chemical Theory and Computation}}} \textbf{\bibinfo{volume}{20}}, \bibinfo{pages}{8152--8174} (\bibinfo{year}{2024}).

\bibitem{Bruneval2015}
\bibinfo{author}{Bruneval, F.}, \bibinfo{author}{Hamed, S.~M.} \& \bibinfo{author}{Neaton, J.~B.}
\newblock \bibinfo{journal}{\bibinfo{title}{A systematic benchmark of the ab initio bethe-salpeter equation approach for low-lying optical excitations of small organic molecules}}.
\newblock {\emph{\JournalTitle{Journal of Chemical Physics}}} \textbf{\bibinfo{volume}{142}}, \bibinfo{pages}{244101}, \doiprefix\url{10.1063/1.4922489} (\bibinfo{year}{2015}).

\bibitem{Rangel2017}
\bibinfo{author}{Rangel, T.}, \bibinfo{author}{Hamed, S.~M.}, \bibinfo{author}{Bruneval, F.} \& \bibinfo{author}{Neaton, J.~B.}
\newblock \bibinfo{journal}{\bibinfo{title}{An assessment of low-lying excitation energies and triplet instabilities of organic molecules with an ab initio bethe-salpeter equation approach and the tamm-dancoff approximation}}.
\newblock {\emph{\JournalTitle{Journal of Chemical Physics}}} \textbf{\bibinfo{volume}{146}}, \bibinfo{pages}{194108}, \doiprefix\url{10.1063/1.4983126} (\bibinfo{year}{2017}).

\bibitem{jacquemin2017benchmark}
\bibinfo{author}{Jacquemin, D.}, \bibinfo{author}{Duchemin, I.}, \bibinfo{author}{Blondel, A.} \& \bibinfo{author}{Blase, X.}
\newblock \bibinfo{journal}{\bibinfo{title}{Benchmark of bethe-salpeter for triplet excited-states}}.
\newblock {\emph{\JournalTitle{Journal of Chemical Theory and Computation}}} \textbf{\bibinfo{volume}{13}}, \bibinfo{pages}{767--783} (\bibinfo{year}{2017}).

\bibitem{OE62}
\bibinfo{author}{Stuke, A.} \emph{et~al.}
\newblock \bibinfo{journal}{\bibinfo{title}{Atomic structures and orbital energies of 61,489 crystal-forming organic molecules}}.
\newblock {\emph{\JournalTitle{Scientific data}}} \textbf{\bibinfo{volume}{7}}, \bibinfo{pages}{58} (\bibinfo{year}{2020}).

\bibitem{SOTA_dataset}
\bibinfo{author}{Fediai, A.}, \bibinfo{author}{Reiser, P.}, \bibinfo{author}{Pe{\~n}a, J. E.~O.}, \bibinfo{author}{Friederich, P.} \& \bibinfo{author}{Wenzel, W.}
\newblock \bibinfo{journal}{\bibinfo{title}{Accurate gw frontier orbital energies of 134 kilo molecules}}.
\newblock {\emph{\JournalTitle{Scientific Data}}} \textbf{\bibinfo{volume}{10}}, \bibinfo{pages}{581} (\bibinfo{year}{2023}).

\bibitem{PBE0_1}
\bibinfo{author}{Adamo, C.} \& \bibinfo{author}{Barone, V.}
\newblock \bibinfo{journal}{\bibinfo{title}{Toward reliable density functional methods without adjustable parameters: The pbe0 model}}.
\newblock {\emph{\JournalTitle{The Journal of chemical physics}}} \textbf{\bibinfo{volume}{110}}, \bibinfo{pages}{6158--6170} (\bibinfo{year}{1999}).

\bibitem{PBE0_2}
\bibinfo{author}{Perdew, J.~P.}, \bibinfo{author}{Burke, K.} \& \bibinfo{author}{Ernzerhof, M.}
\newblock \bibinfo{journal}{\bibinfo{title}{Generalized gradient approximation made simple}}.
\newblock {\emph{\JournalTitle{Physical review letters}}} \textbf{\bibinfo{volume}{77}}, \bibinfo{pages}{3865} (\bibinfo{year}{1996}).

\bibitem{PBE0_3}
\bibinfo{author}{Becke, A.~D.}
\newblock \bibinfo{journal}{\bibinfo{title}{Density-functional thermochemistry. iii. the role of exact exchange}}.
\newblock {\emph{\JournalTitle{The Journal of chemical physics}}} \textbf{\bibinfo{volume}{98}}, \bibinfo{pages}{5648--5652} (\bibinfo{year}{1993}).

\bibitem{Forster2020b}
\bibinfo{author}{F{\"o}rster, A.} \& \bibinfo{author}{Visscher, L.}
\newblock \bibinfo{journal}{\bibinfo{title}{Low-order scaling g0w0 by pair atomic density fitting}}.
\newblock {\emph{\JournalTitle{Journal of Chemical Theory and Computation}}} \textbf{\bibinfo{volume}{16}}, \bibinfo{pages}{7381--7399}, \doiprefix\url{10.1021/acs.jctc.0c00693} (\bibinfo{year}{2020}).

\bibitem{westermayr2021physically}
\bibinfo{author}{Westermayr, J.} \& \bibinfo{author}{Maurer, R.~J.}
\newblock \bibinfo{journal}{\bibinfo{title}{Physically inspired deep learning of molecular excitations and photoemission spectra}}.
\newblock {\emph{\JournalTitle{Chemical Science}}} \textbf{\bibinfo{volume}{12}}, \bibinfo{pages}{10755--10764} (\bibinfo{year}{2021}).

\bibitem{DeltaLearning}
\bibinfo{author}{Ramakrishnan, R.}, \bibinfo{author}{Dral, P.~O.}, \bibinfo{author}{Rupp, M.} \& \bibinfo{author}{Von~Lilienfeld, O.~A.}
\newblock \bibinfo{journal}{\bibinfo{title}{Big data meets quantum chemistry approximations: the $\delta$-machine learning approach}}.
\newblock {\emph{\JournalTitle{Journal of chemical theory and computation}}} \textbf{\bibinfo{volume}{11}}, \bibinfo{pages}{2087--2096} (\bibinfo{year}{2015}).

\bibitem{NNsNeedLotsData_1}
\bibinfo{author}{Golestaneh, P.}, \bibinfo{author}{Taheri, M.} \& \bibinfo{author}{Lederer, J.}
\newblock \bibinfo{journal}{\bibinfo{title}{How many samples are needed to train a deep neural network?}}
\newblock {\emph{\JournalTitle{arXiv preprint arXiv:2405.16696}}}  (\bibinfo{year}{2024}).

\bibitem{NNsNeedLotsData_2}
\bibinfo{author}{Cheng, Y.}, \bibinfo{author}{Petrides, K.~V.} \& \bibinfo{author}{Li, J.}
\newblock \bibinfo{journal}{\bibinfo{title}{Estimating the minimum sample size for neural network model fitting—a monte carlo simulation study}}.
\newblock {\emph{\JournalTitle{Behavioral Sciences}}} \textbf{\bibinfo{volume}{15}}, \bibinfo{pages}{211} (\bibinfo{year}{2025}).

\bibitem{NNsNeedLotsData_3}
\bibinfo{author}{Alwosheel, A.}, \bibinfo{author}{Van~Cranenburgh, S.} \& \bibinfo{author}{Chorus, C.~G.}
\newblock \bibinfo{journal}{\bibinfo{title}{Is your dataset big enough? sample size requirements when using artificial neural networks for discrete choice analysis}}.
\newblock {\emph{\JournalTitle{Journal of choice modelling}}} \textbf{\bibinfo{volume}{28}}, \bibinfo{pages}{167--182} (\bibinfo{year}{2018}).

\bibitem{QM9_1}
\bibinfo{author}{Ruddigkeit, L.}, \bibinfo{author}{Van~Deursen, R.}, \bibinfo{author}{Blum, L.~C.} \& \bibinfo{author}{Reymond, J.-L.}
\newblock \bibinfo{journal}{\bibinfo{title}{Enumeration of 166 billion organic small molecules in the chemical universe database gdb-17}}.
\newblock {\emph{\JournalTitle{Journal of chemical information and modeling}}} \textbf{\bibinfo{volume}{52}}, \bibinfo{pages}{2864--2875} (\bibinfo{year}{2012}).

\bibitem{QM9_2}
\bibinfo{author}{Ramakrishnan, R.}, \bibinfo{author}{Dral, P.~O.}, \bibinfo{author}{Rupp, M.} \& \bibinfo{author}{Von~Lilienfeld, O.~A.}
\newblock \bibinfo{journal}{\bibinfo{title}{Quantum chemistry structures and properties of 134 kilo molecules}}.
\newblock {\emph{\JournalTitle{Scientific data}}} \textbf{\bibinfo{volume}{1}}, \bibinfo{pages}{1--7} (\bibinfo{year}{2014}).

\bibitem{SOTA_model}
\bibinfo{author}{Fediai, A.}, \bibinfo{author}{Reiser, P.}, \bibinfo{author}{Pe{\~n}a, J. E.~O.}, \bibinfo{author}{Wenzel, W.} \& \bibinfo{author}{Friederich, P.}
\newblock \bibinfo{journal}{\bibinfo{title}{Interpretable delta-learning of gw quasiparticle energies from gga-dft}}.
\newblock {\emph{\JournalTitle{Machine Learning: Science and Technology}}} \textbf{\bibinfo{volume}{4}}, \bibinfo{pages}{035045} (\bibinfo{year}{2023}).

\bibitem{SchNet_1}
\bibinfo{author}{Sch{\"u}tt, K.~T.}, \bibinfo{author}{Sauceda, H.~E.}, \bibinfo{author}{Kindermans, P.-J.}, \bibinfo{author}{Tkatchenko, A.} \& \bibinfo{author}{M{\"u}ller, K.-R.}
\newblock \bibinfo{journal}{\bibinfo{title}{Schnet--a deep learning architecture for molecules and materials}}.
\newblock {\emph{\JournalTitle{The Journal of chemical physics}}} \textbf{\bibinfo{volume}{148}} (\bibinfo{year}{2018}).

\bibitem{SchNet_2}
\bibinfo{author}{Sch{\"u}tt, K.} \emph{et~al.}
\newblock \bibinfo{journal}{\bibinfo{title}{Schnet: A continuous-filter convolutional neural network for modeling quantum interactions}}.
\newblock {\emph{\JournalTitle{Advances in neural information processing systems}}} \textbf{\bibinfo{volume}{30}} (\bibinfo{year}{2017}).

\bibitem{DimeNetPlusPlus}
\bibinfo{author}{Gasteiger, J.}, \bibinfo{author}{Giri, S.}, \bibinfo{author}{Margraf, J.~T.} \& \bibinfo{author}{G{\"u}nnemann, S.}
\newblock \bibinfo{journal}{\bibinfo{title}{Fast and uncertainty-aware directional message passing for non-equilibrium molecules}}.
\newblock {\emph{\JournalTitle{arXiv preprint arXiv:2011.14115}}}  (\bibinfo{year}{2020}).

\bibitem{Faleev2004}
\bibinfo{author}{Faleev, S.~V.}, \bibinfo{author}{van Schilfgaarde, M.} \& \bibinfo{author}{Kotani, T.}
\newblock \bibinfo{journal}{\bibinfo{title}{All-electron self-consistent gw approximation: Application to si, mno, and nio}}.
\newblock {\emph{\JournalTitle{Physical Review Letters}}} \textbf{\bibinfo{volume}{93}}, \bibinfo{pages}{126406}, \doiprefix\url{10.1103/PhysRevLett.93.126406} (\bibinfo{year}{2004}).

\bibitem{qsGW_1}
\bibinfo{author}{van Schilfgaarde, M.}, \bibinfo{author}{Kotani, T.} \& \bibinfo{author}{Faleev, S.}
\newblock \bibinfo{journal}{\bibinfo{title}{Quasiparticle self-consistent gw theory}}.
\newblock {\emph{\JournalTitle{Physical review letters}}} \textbf{\bibinfo{volume}{96}}, \bibinfo{pages}{226402} (\bibinfo{year}{2006}).

\bibitem{qsGW_2}
\bibinfo{author}{Kotani, T.}, \bibinfo{author}{Van~Schilfgaarde, M.} \& \bibinfo{author}{Faleev, S.~V.}
\newblock \bibinfo{journal}{\bibinfo{title}{Quasiparticle self-consistent gw method: A basis for the independent-particle approximation}}.
\newblock {\emph{\JournalTitle{Physical Review B -- Condensed Matter and Materials Physics}}} \textbf{\bibinfo{volume}{76}}, \bibinfo{pages}{165106} (\bibinfo{year}{2007}).

\bibitem{bruneval_springer2014}
\bibinfo{author}{Bruneval, F.} \& \bibinfo{author}{Gatti, M.}
\newblock \bibinfo{title}{Quasiparticle self-consistent $\ensuremath{GW}$ method for the spectral properties of complex materials}.
\newblock In \bibinfo{editor}{Di~Valentin, C.}, \bibinfo{editor}{Botti, S.} \& \bibinfo{editor}{Cococcioni, M.} (eds.) \emph{\bibinfo{booktitle}{First Principles Approaches to Spectroscopic Properties of Complex Materials}}, vol. \bibinfo{volume}{347} of \emph{\bibinfo{series}{Topics in Current Chemistry}}, \bibinfo{pages}{99--136}, \doiprefix\url{10.1007/128\_2013\_460} (\bibinfo{publisher}{Springer Berlin Heidelberg}, \bibinfo{year}{2014}).

\bibitem{blase2011first}
\bibinfo{author}{Blase, X.}, \bibinfo{author}{Attaccalite, C.} \& \bibinfo{author}{Olevano, V.}
\newblock \bibinfo{journal}{\bibinfo{title}{First-principles {GW} calculations for fullerenes, porphyrins, phthalocyanine, and other molecules of interest for organic photovoltaic applications}}.
\newblock {\emph{\JournalTitle{Physical Review B}}} \textbf{\bibinfo{volume}{83}}, \bibinfo{pages}{115103} (\bibinfo{year}{2011}).

\bibitem{faber2011first}
\bibinfo{author}{Faber, C.}, \bibinfo{author}{Attaccalite, C.}, \bibinfo{author}{Olevano, V.}, \bibinfo{author}{Runge, E.} \& \bibinfo{author}{Blase, X.}
\newblock \bibinfo{journal}{\bibinfo{title}{First-principles gw calculations for dna and rna nucleobases}}.
\newblock {\emph{\JournalTitle{Physical Review B-Condensed Matter and Materials Physics}}} \textbf{\bibinfo{volume}{83}}, \bibinfo{pages}{115123} (\bibinfo{year}{2011}).

\bibitem{jacquemin2015benchmarking}
\bibinfo{author}{Jacquemin, D.}, \bibinfo{author}{Duchemin, I.} \& \bibinfo{author}{Blase, X.}
\newblock \bibinfo{journal}{\bibinfo{title}{Benchmarking the bethe--salpeter formalism on a standard organic molecular set}}.
\newblock {\emph{\JournalTitle{Journal of Chemical Theory and Computation}}} \textbf{\bibinfo{volume}{11}}, \bibinfo{pages}{3290--3304} (\bibinfo{year}{2015}).

\bibitem{blase2016gw}
\bibinfo{author}{Blase, X.}, \bibinfo{author}{Boulanger, P.}, \bibinfo{author}{Bruneval, F.}, \bibinfo{author}{Fernandez-Serra, M.} \& \bibinfo{author}{Duchemin, I.}
\newblock \bibinfo{journal}{\bibinfo{title}{Gw and bethe-salpeter study of small water clusters}}.
\newblock {\emph{\JournalTitle{The Journal of Chemical Physics}}} \textbf{\bibinfo{volume}{144}} (\bibinfo{year}{2016}).

\bibitem{rangel2016evaluating}
\bibinfo{author}{Rangel, T.}, \bibinfo{author}{Hamed, S.~M.}, \bibinfo{author}{Bruneval, F.} \& \bibinfo{author}{Neaton, J.~B.}
\newblock \bibinfo{journal}{\bibinfo{title}{Evaluating the {GW} approximation with {CCSD(T)} for charged excitations across the oligoacenes}}.
\newblock {\emph{\JournalTitle{Journal of Chemical Theory and Computation}}} \textbf{\bibinfo{volume}{12}}, \bibinfo{pages}{2834--2842} (\bibinfo{year}{2016}).

\bibitem{faber2013many}
\bibinfo{author}{Faber, C.}, \bibinfo{author}{Boulanger, P.}, \bibinfo{author}{Duchemin, I.}, \bibinfo{author}{Attaccalite, C.} \& \bibinfo{author}{Blase, X.}
\newblock \bibinfo{journal}{\bibinfo{title}{Many-body green's function gw and bethe-salpeter study of the optical excitations in a paradigmatic model dipeptide}}.
\newblock {\emph{\JournalTitle{The Journal of Chemical Physics}}} \textbf{\bibinfo{volume}{139}} (\bibinfo{year}{2013}).

\bibitem{Schreiber2008}
\bibinfo{author}{Schreiber, M.}, \bibinfo{author}{Silva-Junior, M.~R.}, \bibinfo{author}{Sauer, S.~P.} \& \bibinfo{author}{Thiel, W.}
\newblock \bibinfo{journal}{\bibinfo{title}{Benchmarks for electronically excited states: Caspt2, cc2, ccsd, and cc3}}.
\newblock {\emph{\JournalTitle{Journal of Chemical Physics}}} \textbf{\bibinfo{volume}{128}}, \doiprefix\url{10.1063/1.2889385} (\bibinfo{year}{2008}).

\bibitem{hashemi2021assessment}
\bibinfo{author}{Hashemi, Z.} \& \bibinfo{author}{Leppert, L.}
\newblock \bibinfo{journal}{\bibinfo{title}{Assessment of the \emph{ab initio} bethe--salpeter equation approach for the low-lying excitation energies of bacteriochlorophylls and chlorophylls}}.
\newblock {\emph{\JournalTitle{The Journal of Physical Chemistry A}}} \textbf{\bibinfo{volume}{125}}, \bibinfo{pages}{2163--2172} (\bibinfo{year}{2021}).

\bibitem{Kshirsagar2023}
\bibinfo{author}{Kshirsagar, A.~R.} \& \bibinfo{author}{Poloni, R.}
\newblock \bibinfo{journal}{\bibinfo{title}{Assessing the role of the kohn-sham density in the calculation of the low-lying bethe-salpeter excitation energies}}.
\newblock {\emph{\JournalTitle{Journal of Physical Chemistry A}}} \textbf{\bibinfo{volume}{127}}, \bibinfo{pages}{2618--2627}, \doiprefix\url{10.1021/acs.jpca.2c07526} (\bibinfo{year}{2023}).

\bibitem{LowOrderScaling_qsGW}
\bibinfo{author}{F{\"o}rster, A.} \& \bibinfo{author}{Visscher, L.}
\newblock \bibinfo{journal}{\bibinfo{title}{Low-order scaling quasiparticle self-consistent gw for molecules}}.
\newblock {\emph{\JournalTitle{Frontiers in Chemistry}}} \textbf{\bibinfo{volume}{9}}, \bibinfo{pages}{736591} (\bibinfo{year}{2021}).

\bibitem{GWBSE_accurate}
\bibinfo{author}{Bruneval, F.} \& \bibinfo{author}{Marques, M. A.~L.}
\newblock \bibinfo{journal}{\bibinfo{title}{Benchmarking the starting points of the {GW} approximation for molecules}}.
\newblock {\emph{\JournalTitle{Journal of Chemical Theory and Computation}}} \textbf{\bibinfo{volume}{9}}, \bibinfo{pages}{324--329} (\bibinfo{year}{2013}).

\bibitem{Forster2025}
\bibinfo{author}{F{\"o}rster, A.}
\newblock \bibinfo{journal}{\bibinfo{title}{Beyond quasi-particle self-consistent {GW} for molecules with vertex corrections}}.
\newblock {\emph{\JournalTitle{Journal of Chemical Theory and Computation}}} \textbf{\bibinfo{volume}{21}}, \bibinfo{pages}{1709--1721}, \doiprefix\url{10.1021/acs.jctc.4c01639} (\bibinfo{year}{2025}).

\bibitem{richard2016accurate}
\bibinfo{author}{Richard, R.~M.} \emph{et~al.}
\newblock \bibinfo{journal}{\bibinfo{title}{Accurate ionization potentials and electron affinities of acceptor molecules {I}. reference data at the {CCSD(T)} complete basis set limit}}.
\newblock {\emph{\JournalTitle{Journal of Chemical Theory and Computation}}} \textbf{\bibinfo{volume}{12}}, \bibinfo{pages}{595--604} (\bibinfo{year}{2016}).

\bibitem{AMS}
\bibinfo{author}{Baerends, E.~J.} \emph{et~al.}
\newblock \bibinfo{journal}{\bibinfo{title}{The amsterdam modeling suite}}.
\newblock {\emph{\JournalTitle{The Journal of Chemical Physics}}} \textbf{\bibinfo{volume}{162}} (\bibinfo{year}{2025}).

\bibitem{PLAMS}
\bibinfo{author}{{SCM}}.
\newblock \bibinfo{title}{Plams: Python library for automating molecular simulation}.
\newblock \bibinfo{howpublished}{\url{https://www.scm.com} and \url{https://github.com/SCM-NV/PLAMS}} (\bibinfo{year}{2025}).
\newblock \bibinfo{note}{Accessed 2025}.

\bibitem{BH}
\bibinfo{author}{Becke, A.~D.}
\newblock \bibinfo{journal}{\bibinfo{title}{A new mixing of hartree-fock and local density-functional theories}}.
\newblock {\emph{\JournalTitle{Journal of chemical Physics}}} \textbf{\bibinfo{volume}{98}}, \bibinfo{pages}{1372--1377} (\bibinfo{year}{1993}).

\bibitem{LYP}
\bibinfo{author}{Lee, C.}, \bibinfo{author}{Yang, W.} \& \bibinfo{author}{Parr, R.~G.}
\newblock \bibinfo{journal}{\bibinfo{title}{Development of the colle-salvetti correlation-energy formula into a functional of the electron density}}.
\newblock {\emph{\JournalTitle{Physical review B}}} \textbf{\bibinfo{volume}{37}}, \bibinfo{pages}{785} (\bibinfo{year}{1988}).

\bibitem{Lehtola2018}
\bibinfo{author}{Lehtola, S.}, \bibinfo{author}{Steigemann, C.}, \bibinfo{author}{Oliveira, M.~J.} \& \bibinfo{author}{Marques, M.}
\newblock \bibinfo{journal}{\bibinfo{title}{Recent developments in libxc — a comprehensive library of functionals for density functional theory}}.
\newblock {\emph{\JournalTitle{SoftwareX}}} \textbf{\bibinfo{volume}{7}}, \bibinfo{pages}{1--5}, \doiprefix\url{10.1016/j.softx.2017.11.002} (\bibinfo{year}{2018}).

\bibitem{GW100_Slater}
\bibinfo{author}{Forster, A.} \& \bibinfo{author}{Visscher, L.}
\newblock \bibinfo{journal}{\bibinfo{title}{Gw100: A slater-type orbital perspective}}.
\newblock {\emph{\JournalTitle{Journal of chemical theory and computation}}} \textbf{\bibinfo{volume}{17}}, \bibinfo{pages}{5080--5097} (\bibinfo{year}{2021}).

\bibitem{Vidberg1977}
\bibinfo{author}{Vidberg, H.~J.} \& \bibinfo{author}{Serene, J.~W.}
\newblock \bibinfo{journal}{\bibinfo{title}{Solving the eliashberg equations by means of n-point padé approximants}}.
\newblock {\emph{\JournalTitle{Journal of Low Temperature Physics}}} \textbf{\bibinfo{volume}{29}}, \bibinfo{pages}{179--192}, \doiprefix\url{10.1007/BF00655090} (\bibinfo{year}{1977}).

\bibitem{Pulay1980}
\bibinfo{author}{Pulay, P.}
\newblock \bibinfo{journal}{\bibinfo{title}{Convergence acceleration of iterative sequences. the case of scf iteration}}.
\newblock {\emph{\JournalTitle{Chemical Physics Letters}}} \textbf{\bibinfo{volume}{73}}, \bibinfo{pages}{393--398}, \doiprefix\url{10.1016/0009-2614(80)80396-4} (\bibinfo{year}{1980}).

\bibitem{Veril2018}
\bibinfo{author}{Véril, M.}, \bibinfo{author}{Romaniello, P.}, \bibinfo{author}{Berger, J.~A.} \& \bibinfo{author}{Loos, P.~F.}
\newblock \bibinfo{journal}{\bibinfo{title}{Unphysical discontinuities in gw methods}}.
\newblock {\emph{\JournalTitle{Journal of Chemical Theory and Computation}}} \textbf{\bibinfo{volume}{14}}, \bibinfo{pages}{5220--5228}, \doiprefix\url{10.1021/acs.jctc.8b00745} (\bibinfo{year}{2018}).

\bibitem{Spadetto2023}
\bibinfo{author}{Spadetto, E.}, \bibinfo{author}{Philipsen, P. H.~T.}, \bibinfo{author}{Förster, A.} \& \bibinfo{author}{Visscher, L.}
\newblock \bibinfo{journal}{\bibinfo{title}{Toward pair atomic density fitting for correlation energies with benchmark accuracy}}.
\newblock {\emph{\JournalTitle{Journal of Chemical Theory and Computation}}} \textbf{\bibinfo{volume}{19}}, \bibinfo{pages}{1499--1516}, \doiprefix\url{10.1021/acs.jctc.2c01201} (\bibinfo{year}{2023}).

\bibitem{Forster2020}
\bibinfo{author}{Förster, A.}, \bibinfo{author}{Franchini, M.}, \bibinfo{author}{van Lenthe, E.} \& \bibinfo{author}{Visscher, L.}
\newblock \bibinfo{journal}{\bibinfo{title}{A quadratic pair atomic resolution of the identity based sos-ao-mp2 algorithm using slater type orbitals}}.
\newblock {\emph{\JournalTitle{Journal of Chemical Theory and Computation}}} \textbf{\bibinfo{volume}{16}}, \bibinfo{pages}{875 -- 891}, \doiprefix\url{https://doi.org/10.1021/acs.jctc.9b00854} (\bibinfo{year}{2020}).

\bibitem{Loos2021}
\bibinfo{author}{Loos, P.~F.}, \bibinfo{author}{Comin, M.}, \bibinfo{author}{Blase, X.} \& \bibinfo{author}{Jacquemin, D.}
\newblock \bibinfo{journal}{\bibinfo{title}{Reference energies for intramolecular charge-transfer excitations}}.
\newblock {\emph{\JournalTitle{Journal of Chemical Theory and Computation}}} \textbf{\bibinfo{volume}{17}}, \bibinfo{pages}{3666--3686}, \doiprefix\url{10.1021/acs.jctc.1c00226} (\bibinfo{year}{2021}).

\bibitem{VanSetten2013}
\bibinfo{author}{Setten, M. J.~V.}, \bibinfo{author}{Weigend, F.} \& \bibinfo{author}{Evers, F.}
\newblock \bibinfo{journal}{\bibinfo{title}{The gw-method for quantum chemistry applications: Theory and implementation}}.
\newblock {\emph{\JournalTitle{Journal of Chemical Theory and Computation}}} \textbf{\bibinfo{volume}{9}}, \bibinfo{pages}{232--246}, \doiprefix\url{10.1021/ct300648t} (\bibinfo{year}{2013}).

\bibitem{Golze2018}
\bibinfo{author}{Golze, D.}, \bibinfo{author}{Wilhelm, J.}, \bibinfo{author}{Setten, M. J.~V.} \& \bibinfo{author}{Rinke, P.}
\newblock \bibinfo{journal}{\bibinfo{title}{Core-level binding energies from gw: An efficient full-frequency approach within a localized basis}}.
\newblock {\emph{\JournalTitle{Journal of Chemical Theory and Computation}}} \textbf{\bibinfo{volume}{14}}, \bibinfo{pages}{4856--4869}, \doiprefix\url{10.1021/acs.jctc.8b00458} (\bibinfo{year}{2018}).

\bibitem{Golze2020}
\bibinfo{author}{Golze, D.}, \bibinfo{author}{Keller, L.} \& \bibinfo{author}{Rinke, P.}
\newblock \bibinfo{journal}{\bibinfo{title}{Accurate absolute and relative core-level binding energies from gw}}.
\newblock {\emph{\JournalTitle{The Journal of Physical Chemistry Letters}}} \textbf{\bibinfo{volume}{11}}, \bibinfo{pages}{1840--1847}, \doiprefix\url{10.1021/acs.jpclett.9b03423} (\bibinfo{year}{2020}).

\bibitem{slow_basis_set_convergence_1}
\bibinfo{author}{Harsha, G.}, \bibinfo{author}{Abraham, V.} \& \bibinfo{author}{Zgid, D.}
\newblock \bibinfo{journal}{\bibinfo{title}{Challenges with relativistic gw calculations in solids and molecules}}.
\newblock {\emph{\JournalTitle{Faraday Discussions}}} \textbf{\bibinfo{volume}{254}}, \bibinfo{pages}{216--238} (\bibinfo{year}{2024}).

\bibitem{GWExtrpol_1}
\bibinfo{author}{Li, J.}, \bibinfo{author}{Jin, Y.}, \bibinfo{author}{Rinke, P.}, \bibinfo{author}{Yang, W.} \& \bibinfo{author}{Golze, D.}
\newblock \bibinfo{journal}{\bibinfo{title}{Benchmark of gw methods for core-level binding energies}}.
\newblock {\emph{\JournalTitle{Journal of Chemical Theory and Computation}}} \textbf{\bibinfo{volume}{18}}, \bibinfo{pages}{7570--7585} (\bibinfo{year}{2022}).

\bibitem{GWExtrpol_2}
\bibinfo{author}{van Setten, M.~J.}, \bibinfo{author}{Costa, R.}, \bibinfo{author}{Vines, F.} \& \bibinfo{author}{Illas, F.}
\newblock \bibinfo{journal}{\bibinfo{title}{Assessing gw approaches for predicting core level binding energies}}.
\newblock {\emph{\JournalTitle{Journal of Chemical Theory and Computation}}} \textbf{\bibinfo{volume}{14}}, \bibinfo{pages}{877--883} (\bibinfo{year}{2018}).

\bibitem{GWExtrpol_3}
\bibinfo{author}{Van~Setten, M.~J.} \emph{et~al.}
\newblock \bibinfo{journal}{\bibinfo{title}{Gw 100: Benchmarking g 0 w 0 for molecular systems}}.
\newblock {\emph{\JournalTitle{Journal of chemical theory and computation}}} \textbf{\bibinfo{volume}{11}}, \bibinfo{pages}{5665--5687} (\bibinfo{year}{2015}).

\bibitem{QP_CBS_Extrpol}
\bibinfo{author}{Baum, D.}, \bibinfo{author}{Visscher, L.} \& \bibinfo{author}{Förster, A.}
\newblock \bibinfo{journal}{\bibinfo{title}{Predicting complete basis set limit quasiparticle energies from triple-$\zeta$ calculations}}.
\newblock {\emph{\JournalTitle{arXiv preprint}}} \doiprefix\url{10.48550/arXiv.2511.22462} (\bibinfo{year}{2025}).
\newblock \eprint{2511.22462}.

\bibitem{TwoPointExtrpol_1}
\bibinfo{author}{Helgaker, T.}, \bibinfo{author}{Klopper, W.}, \bibinfo{author}{Koch, H.} \& \bibinfo{author}{Noga, J.}
\newblock \bibinfo{journal}{\bibinfo{title}{Basis-set convergence of correlated calculations on water}}.
\newblock {\emph{\JournalTitle{The Journal of chemical physics}}} \textbf{\bibinfo{volume}{106}}, \bibinfo{pages}{9639--9646} (\bibinfo{year}{1997}).

\bibitem{TwoPointExtrpol_2}
\bibinfo{author}{Halkier, A.} \emph{et~al.}
\newblock \bibinfo{journal}{\bibinfo{title}{Basis-set convergence in correlated calculations on ne, n2, and h2o}}.
\newblock {\emph{\JournalTitle{Chemical Physics Letters}}} \textbf{\bibinfo{volume}{286}}, \bibinfo{pages}{243--252} (\bibinfo{year}{1998}).

\bibitem{Bruneval2020}
\bibinfo{author}{Bruneval, F.}, \bibinfo{author}{Maliyov, I.}, \bibinfo{author}{Lapointe, C.} \& \bibinfo{author}{Marinica, M.-C.}
\newblock \bibinfo{journal}{\bibinfo{title}{Extrapolating unconverged gw energies up to the complete basis set limit with linear regression}}.
\newblock {\emph{\JournalTitle{Journal of Chemical Theory and Computation}}} \textbf{\bibinfo{volume}{16}}, \bibinfo{pages}{4399--4407}, \doiprefix\url{10.1021/acs.jctc.0c00433} (\bibinfo{year}{2020}).

\end{thebibliography}

\section*{Acknowledgements}

We acknowledge the use of supercomputer facilities at SURFsara sponsored by NWO Physical Sciences, with financial support from The Netherlands Organization for Scientific Research (NWO). LV and DB acknowledge funding from Microsoft Research. AF acknowledges funding through a VENI grant from NWO under grant agreement VI.Veni.232.013. We also thank Ansgar Pausch for substantial contributions and suggestions in the initial ideation phase.

\section*{Author Contributions}

D.B. curated the data, carried out the calculations and postprocessed the results. A.F. and D.B. conceived and validated the calculations. D.B., A.F. and L.V. conceived the original idea and designed the study. A.F. and L.V. provided guidance and supervision throughout the project. L.V. acquired funding that supported this work. All authors cowrote and reviewed the manuscript.

\section*{Competing Interests}

The authors declare no competing interests.

\end{document}